\documentclass[a4paper]{jpconf}
\usepackage{graphicx}
\begin{document}
\title{Low temperature thermal resistance for a new design of silver sinter heat exchanger}

\author{J. Pollanen, H. Choi, J.P. Davis, B.T. Rolfs and W.P. Halperin}

\address{Department of Physics and Astronomy, Northwestern University, Evanston, IL 60208, USA}

\ead{j-pollanen@northwestern.edu}

\begin{abstract}
We have developed a novel procedure for constructing high surface area silver sinter heat exchangers.  Our recipe incorporates nylon fibers having a diameter of $\sim 50 ~\mu$m and thin wires of bulk silver in the heat exchanger.  In order to increase the thermal conductance of liquid helium within the heat exchanger, prior to sintering, the nylon fibers are dissolved with an organic acid leaving a network of channels.  In addition, the silver wires reinforce the structural integrity, and reduce the resistance, of the silver sinter.  We have constructed a $^3$He melting curve thermometer (MCT) with this type of heat exchanger and measured the thermal time response of the liquid $^3$He inside the MCT in the temperature range $T \approx 2-150$ mK.  We find a thermal relaxation time of $\sim 490$ s at $T \approx 1$ mK.  We have used scanning electron microscopy (SEM) to characterize the heat exchanger and BET absorption for determination of the specific surface area.
\end{abstract}

\section{Introduction}
Heat exchangers constructed from metallic sinters are well suited, and widely used, to cool liquid helium to low temperatures.  The large surface areas attainable in metallic sinters can compensate for the large Kapitza boundary resistance at the metal-liquid helium interface.  However, depending upon the open volume of the sinter, the thermal resistance within the liquid helium or in the sinter itself can be of the same order as, or larger than, the Kapitza resistance \cite{Pob92}, in particular well below the transition to superfluidity in $^{3}$He.  Our construction procedure for low temperature heat exchangers is designed to reduce these two, non-Kapitza, contributions to the overall thermal resistance by increasing the thermal conductance of the liquid helium and the conductance within the sinter.  This is performed by creating a series of channels in the sinter having a diameter of $\sim 50 ~\mu$m, allowing greater access to the internal surface area of the exchanger and adding bulk silver wires throughout the heat exchanger.  Here we report on our heat exchanger construction procedure and measurements of the low temperature thermal resistance.

\section{Heat Exchanger Construction}
The powder used for our heat exchanger was Grade 440 silver powder purchased from Nanopowder Industries (Israel).  Using BET absorption \cite{Bru38} of nitrogen we measured the specific surface area of the raw powder to be 2.6 $\pm$ 0.1 m$^{2}$/g, which is comparable to, or larger than, values found for other powders typically used in the construction of low temperature heat exchangers \cite{Bus84}.  Note, prior to the BET measurement the powder was sifted through a fine nylon mesh having openings of $\sim 50 ~\mu$m.

\begin{figure}
\begin{minipage}[t]{0.47\linewidth}
\centering
\includegraphics[height=5cm]{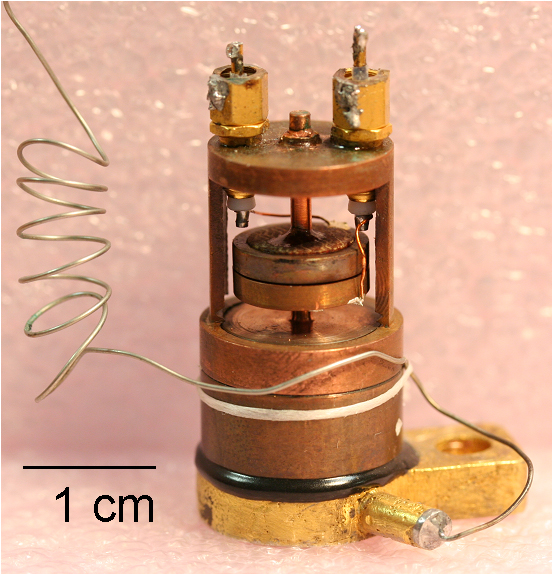}
\caption{\label{fig1}(Color online) $^{3}$He MCT used in the thermal relaxation measurements.}
\end{minipage}
\hspace{0.5cm}
\begin{minipage}[t]{0.47\linewidth}
\centering
\includegraphics[height=5cm]{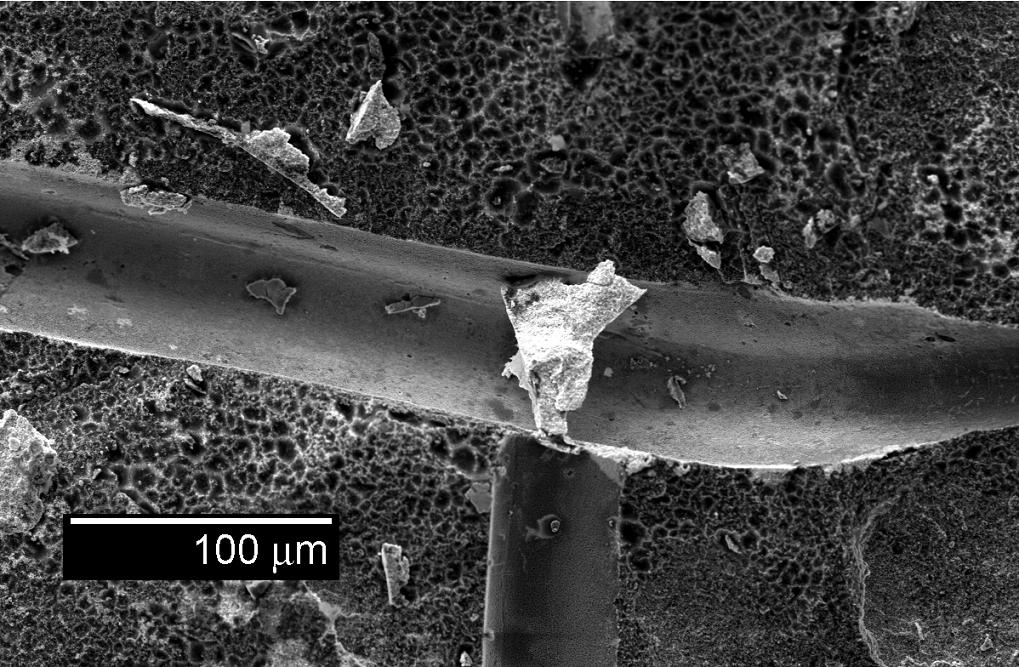}
\caption{\label{fig2}(Color online) SEM image of the sintered silver heat exchanger.  Note the $50$ $\mu$m channels present after dissolving the nylon fibers.}
\end{minipage}
\end{figure}

A mixture of sifted powder, along with $\sim 50 ~\mu$m diameter nylon fibers, and $\sim 170-250 ~\mu$m diameter bulk silver wires, was pressed into the gold plated copper body of a $^{3}$He melting curve thermometer \cite{Ada93,Gre85}, see Fig.~\ref{fig1}.
The nylon fibers and silver wires were cut to $\sim 1-2$ cm lengths to ensure uniform coverage across the 1.1 cm diameter of the heat exchanger.  The mass mix ratio for the three ingredients was \{powder : nylon : wires = 1 : 0.017 : 0.052\}.  This mixture was divided into seven equal charges that were sequentially pressed in the body of the MCT at a constant pressure of $\sim 45$ MPa.  For each charge the pressure was applied for approximately 60 s.  Care was taken to maintain a homogeneous distribution of the ingredients within each charge.  A circular piece of coarse filter paper was placed between the mixture and the surface of the brass press plunger during each charge and was removed between subsequent presses.  This procedure produced a rough pressed powder surface to obtain good contact between the material in consecutive charges.  After seven charges, the final height of the cylindrical heat exchanger was 2 mm.

The nylon fibers were then removed by placing the heat exchanger into a beaker containing 100 cc of formic acid (CH$_{2}$O$_{2}$).  The formic acid penetrates into the pores of the heat exchanger and dissolves the nylon, leaving $\sim 50 ~\mu$m channels.  The acid bath was replaced each day and the exchanger remained immersed for a total of three days after which it was transferred to a 500 cc methanol bath to solvent exchange the formic acid.  The methanol was replaced each day, for two days after which a similar procedure was followed to solvent exchange the methanol with pure water.  The heat exchanger was allowed to dry and then sintered at 125$^\circ$C for 45 minutes in an evacuated oven that was back-filled with hydrogen gas to a pressure of $\sim 75$ torr.

The final packing fraction of the heat exchanger was measured to be $\sim 56$\% relative to the density of bulk silver.  We performed SEM to characterize the channels in the heat exchanger.  Fig.~\ref{fig2} is a representative SEM image.  We have checked the specific surface area of heat exchangers made by this method with nitrogen BET \cite{Bru38} and find a value of 1.8 $\pm$ 0.1 m$^{2}$/g.  The amount of silver powder in our heat exchanger was 1.59 $\pm$ 0.01 g yielding a total surface area of 2.9 $\pm$ 0.2 m$^{2}$.  

\section{Thermal Relaxation Measurement}
\begin{figure}
\begin{minipage}[t]{0.47\linewidth}
\centering
\includegraphics[height=6cm]{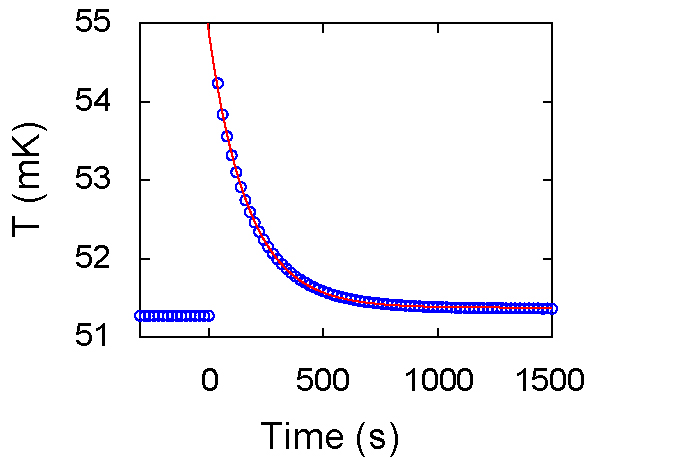}
\caption{\label{fig3}(Color online) Exponential recovery of the temperature of the of the $^{3}$He inside the MCT at $\sim 50$ mK.  The solid red curve is a single exponential fit to the data used to determine the time constant, $\tau$.}
\end{minipage}
\hspace{0.5cm}
\begin{minipage}[t]{0.47\linewidth}
\centering
\includegraphics[height=6cm]{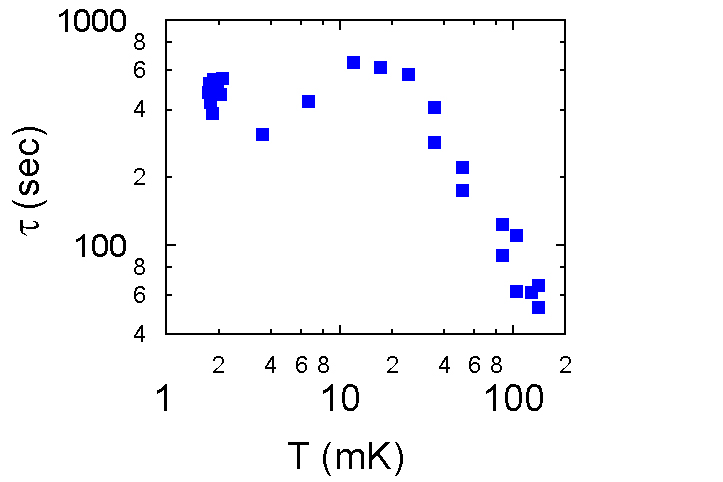}
\caption{\label{fig4}(Color online) Thermal time constant, $\tau$, as a function of temperature for our heat exchanger.  The increase in $\tau$ at low temperatures is due to the increased heat capacity of $^{3}$He upon entering the superfluid state.}
\end{minipage}
\end{figure}
To ascertain the quality of our heat exchanger we have made measurements of the thermal relaxation of the $^3$He inside the MCT as a function of temperature in the range $T \approx 1-100$ mK.  The MCT, which was calibrated with respect to the Greywall temperature scale \cite{Gre86}, monitored the temperature of the $^{3}$He.  The volume of $^{3}$He in the MCT was $0.55 \pm 0.1$ cc.  The MCT was attached to a copper flange connected to a PrNi$_{5}$ nuclear demagnetization refrigerator.  All of the measurements were performed with zero magnetic field at the position of the MCT.  Cooling to $\sim 20$ mK was performed with a dilution refrigerator during which a magnetic field of 8 T was applied to the PrNi$_{5}$ nuclear stage.  Temperatures below this were achieved by adiabatically dimagnetizing the PrNi$_{5}$.  At the lowest temperatures the final demagnetization field was 300 G.  Data were collected by applying heat pulses to the nuclear stage to raise the temperature of the $^{3}$He inside the MCT.  The exponential recovery of the temperature of the $^{3}$He was fit to obtain the recovery time constant, $\tau$, see Fig.~\ref{fig3}.

\section{Results and Discussion}
\begin{figure}
\centerline{\includegraphics[height=7cm]{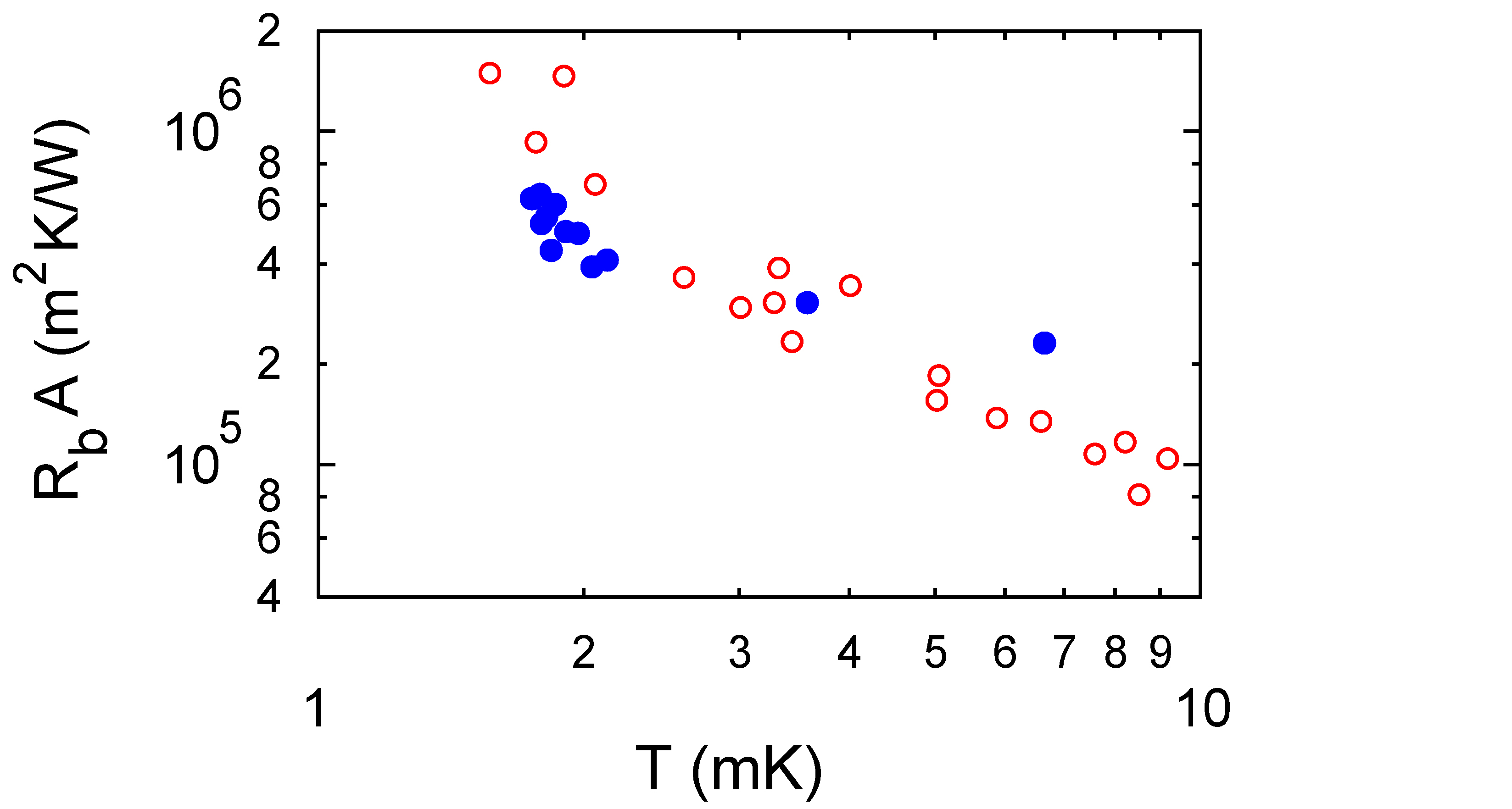}}
\caption{\label{fig5}(Color online) Low temperature values of $R_{b}A$ for our heat exchanger (closed blue cirlces) along with data from Andres \emph{et. al} \cite{And75} (open red circles).}
\end{figure}

In Fig.~\ref{fig4} we present results of the thermal relaxation time constant, $\tau$, for our heat exchanger as a function of temperature.  At temperatures above $\sim 20$ mK we find that $\tau$ is longer than would be expected from a $T^{-3}$ temperature dependence of the Kaptiza boundary resistance \cite{Pob92,And75} and may reflect a reduction in the thermal conductance of the $^{3}$He within the sinter \cite{Cou94}, possibly due to the formation of solid $^{3}$He within the channels.  From our measurements of $\tau$ it is possible to calculate the thermal boundary resistance $R_{b}$ between the $^{3}$He and silver sinter according to the following relation, $R_{b}=\tau / C_{3}$, where $C_{3}$ is the heat capacity of the $^{3}$He inside the MCT \cite{Gre86,Hal90}.  This relation is valid for our experimental configuration because at all temperatures $C_{3} << C_{NS}$, where $C_{NS}$ is the heat capacity of the PrNi$_{5}$ nuclear stage \cite{Kub80}.  In Fig.~\ref{fig5} we present results for the thermal boundary resistance, $R_{b}$, multiplied by the surface area, $A$, of the silver sinter for our heat exchanger along with the results of Andres \emph{et al.} \cite{And75} in the temperature range $T = 1-10$ mK. We note that at 2 mK $R_{b}A$ for our sinter is approximately 2.4 times lower than that reported by Andres \emph{et al.} \cite{And75} and corresponds to $\tau = 490$ s.

\section{Conclusions}
We have measured the low temperature thermal boundary resistance between liquid $^{3}$He and our novel silver sinter heat exchanger inside of a MCT.  At $\sim 2$ mK we find $R_{b}A = 6.28 (\pm 0.01) \times 10^{5}$ m$^{2}$K/W, which is comparable to, or lower than, typical values found for other silver sinters \cite{And75}.  Additionally, our construction method perserves $\sim 70$\% of the specific surface area of the raw powder and could be useful in the construction of compact, efficient low temperature heat exchangers.

\ack
We acknowledge support from the National Science Foundation DMR-0703656 and thank G.R. Pickett for valuable discussions.


\section{References}
\begin{thebibliography}{9}
\bibitem{Pob92} F. Pobell, {\em Matter and Methods at Low Temperatures}, (Springer-Verlag, Berlin Heidelberg) (1992).
\bibitem{Bru38} S. Brunauer, P. H. Emmett and E. Teller, \emph{J. Am. Chem. Soc.} {\bf 60}, 309 (1938).
\bibitem{Bus84} P.A. Busch, S.P. Cheston and D.S. Greywall, \emph{Cryogenics} August 1984, 446 (1984).
\bibitem{Ada93} E.D. Adams, \emph{Rev. Sci. Instrum.} {\bf64}, 601 (1993).
\bibitem{Gre85} D.S. Greywall, \emph{Phys. Rev. B} {\bf31}, 2675 (1985). 
\bibitem{Gre86} D.S. Greywall, \emph{Phys. Rev. B} {\bf33}, 7520 (1986).
\bibitem{And75} K. Andres and W. Sprenger, \emph{Proc. 14th Intl. Conf. Low Temp. Phys.} {\bf1}, 123, ed. M. Krusius and M. Vuorio, (North-Holland Amsterdam) (1975).
\bibitem{Cou94} D.J. Cousins, S.N. Fisher, A.M. Gu\'enault, G.R. Pickett, E.N. Smith and R.P. Turner, \emph{Phys. Rev. Lett.} {\bf73}, 2583 (1994).
\bibitem{Hal90} W.P. Halperin and E. Varoquax, \emph{Helium Three} ed. W.P. Halperin and L.P. Pitaevskii, (Elsevier) (1990).
\bibitem{Kub80} M. Kubota, H. R. Folle, Ch. Buchal, R. M. Mueller and F. Pobell, \emph{Phys. Rev. Lett.} {\bf45}, 1812 (1980).
\end{thebibliography}
\end{document}